# Cognitive Radios: A Survey of Methods for Channel State Prediction.

Ashish Kumar, Lakshay Narula and Surya Pal Singh
Department of Electronics Engineering
Indian Institute of Technology-BHU (Varanasi) – 221005
Email: ashish.kumar.ece10@itbhu.ac.in, lakshay.narula.ece10@itbhu.ac.in, spsingh.ece@itbhu.ac.in

## I. Introduction

### A. Background

With the increasing demand of wireless application, the insufficiency of the electromagnetic radio spectrum is getting more and more serious. Spectrum is regulated by governmental agencies that license their use to individuals or organisations on a long-term basis, normally over huge geographical regions. The Nov. 2002 report of the FCC speaks of *spectrum access in many bands as a more significant problem than the scarcity of spectrum itself, owing mainly to the restrictions imposed by the legacy command-and-control regulations over the potential spectrum users* [1]. Due to this, precious resources go wasted for large-frequency regions are used very sporadically.

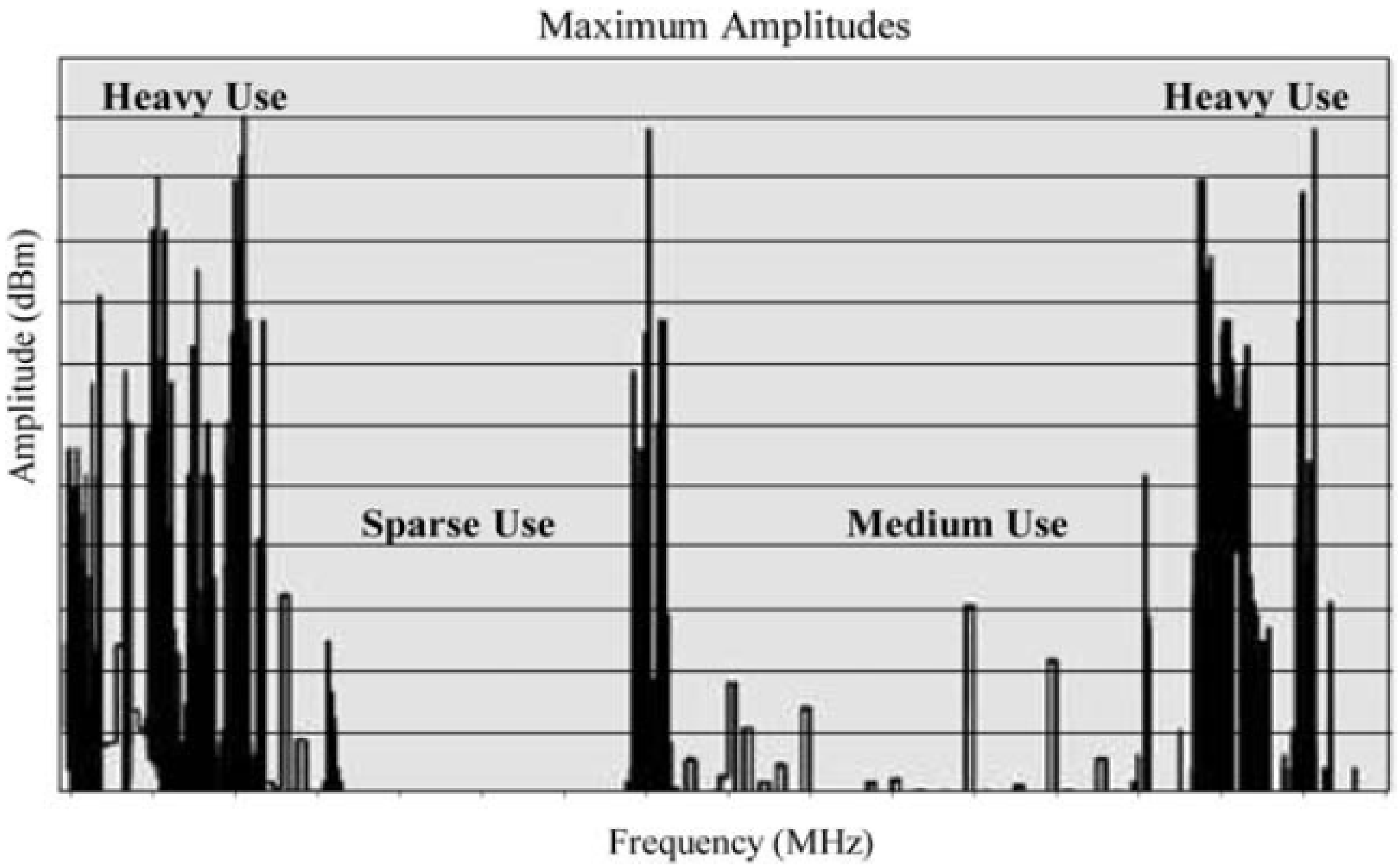

Fig. 1: Frequency use and distribution in Radio Spectrum.

This as a fact has been long established by the findings of many studies on the radio spectrum, which prompted us to think in terms of *spectrum holes.* A spectrum hole has been defined as "a band of frequencies assigned to a primary user, but at a particular time and specific geographic location, the band is not being utilized by that user" [2]. So, basically, our spectrum comprises of frequency bands that are largely unoccupied most of the time (white spaces), some other bands that are underutilized (grey spaces) and heavily utilized bands (black spaces).

Spectrum utilization can be improved significantly by permitting opportunistic access to white and grey spaces at the right place and time. "The vision is to assign appropriate resources to the end users only as long as they are needed for a geographically bounded region, i.e. a personal, local, regional or global cell" [3]. The spectrum access is then organised by the users in the network. *Cognitive radio*, inclusive of the software-defined radio, was proposed by Joseph Mitola *et al.* [4] as the means to promote the efficient use of spectrum by exploiting the existence of spectrum holes.

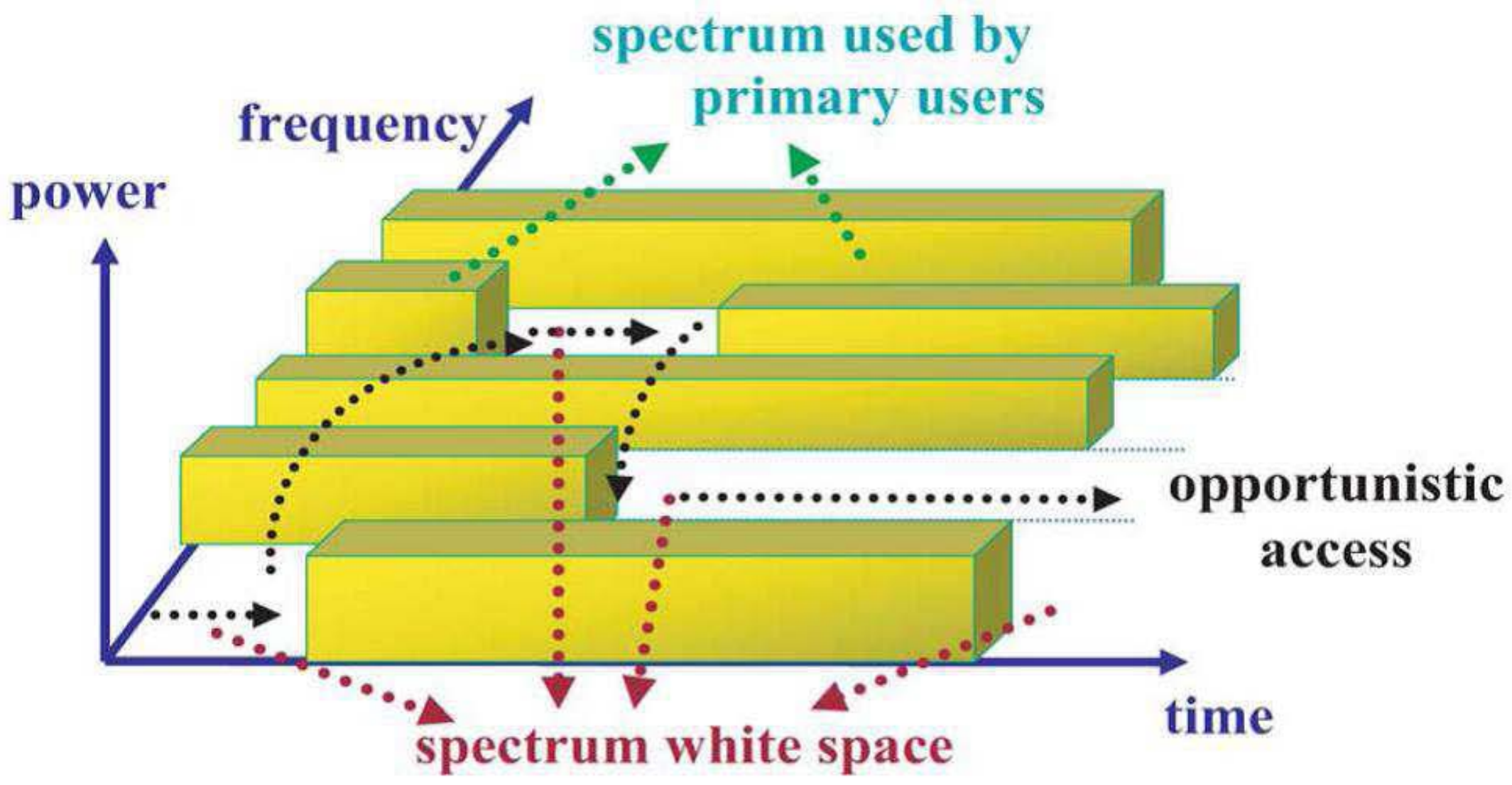

Fig. 2: Opportunistic access of spectrum.

### B. *Cognitive Radio*

Haykin in his paper [5] has defined cognitive radio as an intelligent, reconfigurable wireless communication system that is aware of its surrounding environment (i.e. outside world), and uses the methodology of understanding-by-building to learn from the environment and adapt its internal states to statistical variations in the incoming RF stimuli by making corresponding changes in certain operating parameters (e.g. transmit-power, carrier-frequency, and modulation strategy) in real-time, with two primary objectives in mind:

i. Highly reliable communications whenever and wherever needed.
ii. Efficient utilization of the radio spectrum.

Starting from the transmission demand of its user, the cognitive radio decides about the data rate, the transmission mode and therefore about the bandwidth of the transmission. Afterwards it has to find an appropriate resource for its transmission. This presumes that the cognitive radio knows where it is (self-location), what it is able to do (self-awareness), and where the reachable base stations are. To get more information about possible interferences, it should, for example, be able to detect signals active in adjacent frequency bands and to recognize their transmission standards.

Cognitive radio implements advanced spectrum management through the processes of spectrum reallocation, leasing and spectrum sharing. To implement the network of secondary user (SU) into the system of primary user (PU), it must be ensured that there is no disturbance to the PU system by keeping the transmit power of SU under check and that all signal processing required to avoid disturbances of PUs communication must be implemented in the SUs' system. A successful SUs network needs reliable detection of upcoming PUs' signal within an extremely short interval of time and signalling of current transmission situation in PUs' system to all stations of SUs' system so that they do not use the frequencies occupied by the PUs. This can be done by distributed detection, boosting of the detection results and combining them with the hotspot's access point to an occupancy vector, and distributing the occupancy vector to all base stations in the hotspot [3].

Thus, the next step in the evolution of intelligent transmission devices leads to cognitive radios that may be looked upon as a small part of the physical world using and providing information over very different time scales. This futuristic approach is still in its inchoate stage and it is important that we have a look at an integral cognitive task of our intelligent radio, that of *channel state estimation*, to see how best we can develop this technique to empower our cognitive radio.

### C. Purpose of this Paper

A cognitive radio wireless channel comprises of the transmitter, the physical transmission channel, the receiver in the complex baseband and a feedback channel. The cognitive modules in the transmitter work in tandem with the cognitive modules in the receiver. The processes of radio-scene analysis and channel state estimation are done at the receiver end and the information is conveyed to the transmitter cognitive module through a feedback channel. Such an evaluation of the network performance based on signal processing and machine-learning procedures provides insight into techniques for optimizing future communication protocols and is an important part in the usage of testing and analysing error-control schemes.

This paper studies few methods of channel state prediction using the machine-learning procedures on which channel behaviour models can be built for our cognitive radio wireless network and discusses their performance.

### D. *Organisation of this Paper*

The remaining parts of this paper are organized as follows.

i. Section II discusses the process of channel state prediction: its meaning and the importance of this process in wireless communication.
ii. Section III is devoted to the methods of wireless channel state prediction and analysis of the same.
iii. Section IV concludes the paper and highlights issues that merit attention in the future development of cognitive radio.

## II. Channel State Prediction

### A. *Meaning*

Channel state prediction in cognitive wireless networks involves the analysis of the radio environment and evaluation of certain parameters which can be used to make the best decision for optimal communication. In this context, we can talk of CSI, i.e. Channel State Information which refers to the channel properties that affect the propagation of signal from the transmitter to the receiver.

In cognitive radios, CSI is estimated at the receiver and usually quantized and fed back to the transmitter via feedback channel. The cognitive module at the transmitter end receives the information and based on the history of signal transmitted and received, it trains itself using the machine-learning algorithms. The more comprehensive its test database, the better is it trained to predict the channel behaviour and this allows the cognitive radio to exploit the resources at its disposal.

The CSI acquisition is practically limited by the rate at which the channel conditions change. The available CSI may in general comprise of the instantaneous CSI with some quantization/estimation error and the statistical CSI.

### B. *Need*

Channel selection is one of the most important functions of the cognitive engine in any cognitive radio. To assign a channel to the cognitive radio, it has to be seen that the channel in question is vacant (spectrum sensing will not be discussed) and it should be free for a longer period of time based on the channel history available in database. We are interested in creating a model of our wireless channel so that we can better understand the communication being carried on in it. Thus, predicting the next state of the channel is very important and we will focus on that aspect.

The CSI makes it possible to adapt transmissions to current channel conditions, which is crucial for achieving reliable communication with high data rates. There are many methods that make use of the CSI and empower the cognitive radio to set up a communication link which can adapt the transmitted signal to the channel characteristics and thereby optimize the received signal to achieve low bit error rates.

## III. Methods of Channel State Prediction

Evaluating a mobile wireless network is a challenging problem because the quality of the wireless link varies unpredictably over time and space. To have a better understanding of the behavior of the wireless links and to optimize the protocols at various levels of the network hierarchy, we require cognitive functionalities. The Encyclopaedia of Computer Science [6] discusses a three-point computational view of cognition:

i. Mental states and processes intervene between input stimuli and output responses.
ii. The mental states and processes are described by algorithms.
iii. The mental states and processes lend themselves to scientific investigations.

Moreover, the interdisciplinary study of cognition is concerned with exploring general principles of intelligence through a synthetic methodology termed *learning by understanding* [7]. It is the usage of this technique for predicting and simulating the channel conditions that plays an important role in understanding network protocols and application behaviour. And a Hidden Markov Model (HMM) is the most widely used method to apply machine learning to data that is represented as a sequence of observations over time.

### *Markov Model and HMM*

A Markov model is a powerful abstraction for time series data. For any given system, a Markov model consists of a list of the possible states of that system, the possible transition paths between those states and the rate parameters of those transitions under the assumptions that the transitions to the future states of the process depend only on the present state and conditional distribution to next state given current state does not change over time. However a Markov model fails to capture a common scenario where we reason about the series of states while we cannot observe the states themselves, but some probabilistic function of those states only. In such a case, we can use HMMs to generate the most likely series of states using only some outcome observed at each state.

In a HMM, we have a series of observed outputs available from the set of all possible outputs and we assume the existence of a series of states drawn from the set of all possible states but in this scenario, the values of states are unobserved. A *transition probability matrix* A represents the transition between states which depends only on the previous state and an *emission probability matrix* B encodes the probability of our hidden state generating an output, given the state at that corresponding time. With the *initial state probability* π given, such a HMM can be used to answer the questions that demand the probability of an observed sequence or the most likely series of states to generate the observations.

Generally, the input part of the channel state predictor receives channel history and feeds it to the HMM parameter training module. There, the HMM parameters A, B and π are trained and thereafter, posterior probability (probability for a state to be the next state, conditioned on the previous history) is computed. Finally, the next state would be decided by finding the state with the highest posterior probability.

Now we are in a position to discuss few methods for wireless channel state prediction and modelling that use above concepts.

## A. Markov-based Trace Analysis

Markov-based Trace Analysis (MTA) approach to channel modelling was proposed by Almudena Konrad, Ben Zhao et al. in their paper "A Markov based Channel Model Algorithm for Wireless Networks". It essentially focusses on the design of channel error models, analysing the wireless trace to derive a statistical constant and using this constant to identify the lossy and error-free segments of transmission. The length distribution of the lossy and error-free subtraces is used to effectively characterize the transitions between them and develop a model to accurately represent the original trace. While the earlier works assumed static error statistics, time-varying error statistics can be attended to in this approach for improved accuracy.

So, basically, the MTA algorithm defines two states, *lossy* and *error-free* states and assumes that a frame error trace with non-stationary error properties can be decomposed into them. Lossy states are stationary subsets that contain both error-free and error frames while error-free states contain only correctly transmitted frames. An appropriate window size is used to examine the GSM trace and define stationary trace as one whose error statistics remain relatively constant over time. Showing that in the error trace, error-free bursts are significantly longer than the error bursts, a parameter called *change-of-state constant* C is introduced. C is a design decision which is defined as the mean plus one standard deviation of the length of error bursts of a trace. Trace sections that exhibit stationary properties have been identified by finding error-free bursts of length greater than or equal to C. Thus, a lossy state can be defined as a sequence of 0s and 1s (always started by a 1, where 0 indicates correct frame and 1 indicates corrupted frame), where each run of 0s is not greater than C.

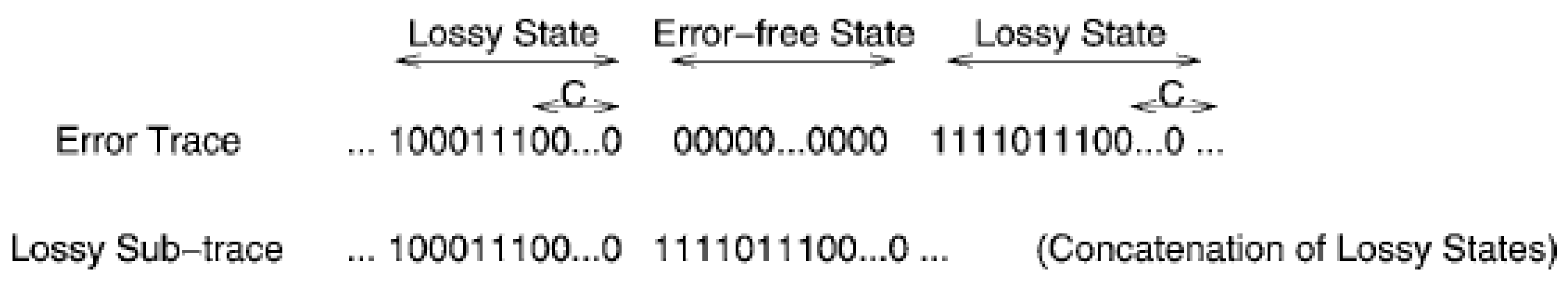

Fig. 3: Separation of a hypothetical error trace into two stationary traces [8].

The lossy states are extracted from the error trace and concatenated to create a lossy subtrace using MTA algorithm. The lossy subtrace is shown to be a stationary random process and hence, can be modelled as a Discrete Time Markov Chain (DTMC) with a certain memory. Then, the MTA algorithm calculates the memory of lossy subtrace without going for undue complexity due to higher order Markov model and determines its transition probabilities.

Finally, the MTA algorithm determines best fitting distribution for the lossy state length process and the error-free state length process by plotting CDF for both the processes. The CDF function is: $F(x) = 1 - e^{-\alpha x}$ where $x$ is error-free or lossy state length and $\alpha$ is a parameter ranging from 0 to 1 whose value for the best approximation to the CDF for GSM trace is chosen. Next, the distribution with the smallest standard error (error between the plots) which signifies a more accurate prediction is chosen.

The performance of MTA model is tested against Gilbert model [9] (a first order DTMC) and a third order Markov model. Knowing the transition probabilities for each, artificial traces are generated from all the three models and compared against the GSM trace. It is seen that the error length statistics (in terms of mean, standard deviation and the maximum length) of the MTA artificial trace bear strongest resemblance to the GSM trace with lowest standard error. Thus, the value of $\alpha$ in the CDF analysis that gives the most accurate modelling of GSM trace is also the one which more accurately portrays the GSM trace in the retrace analysis.

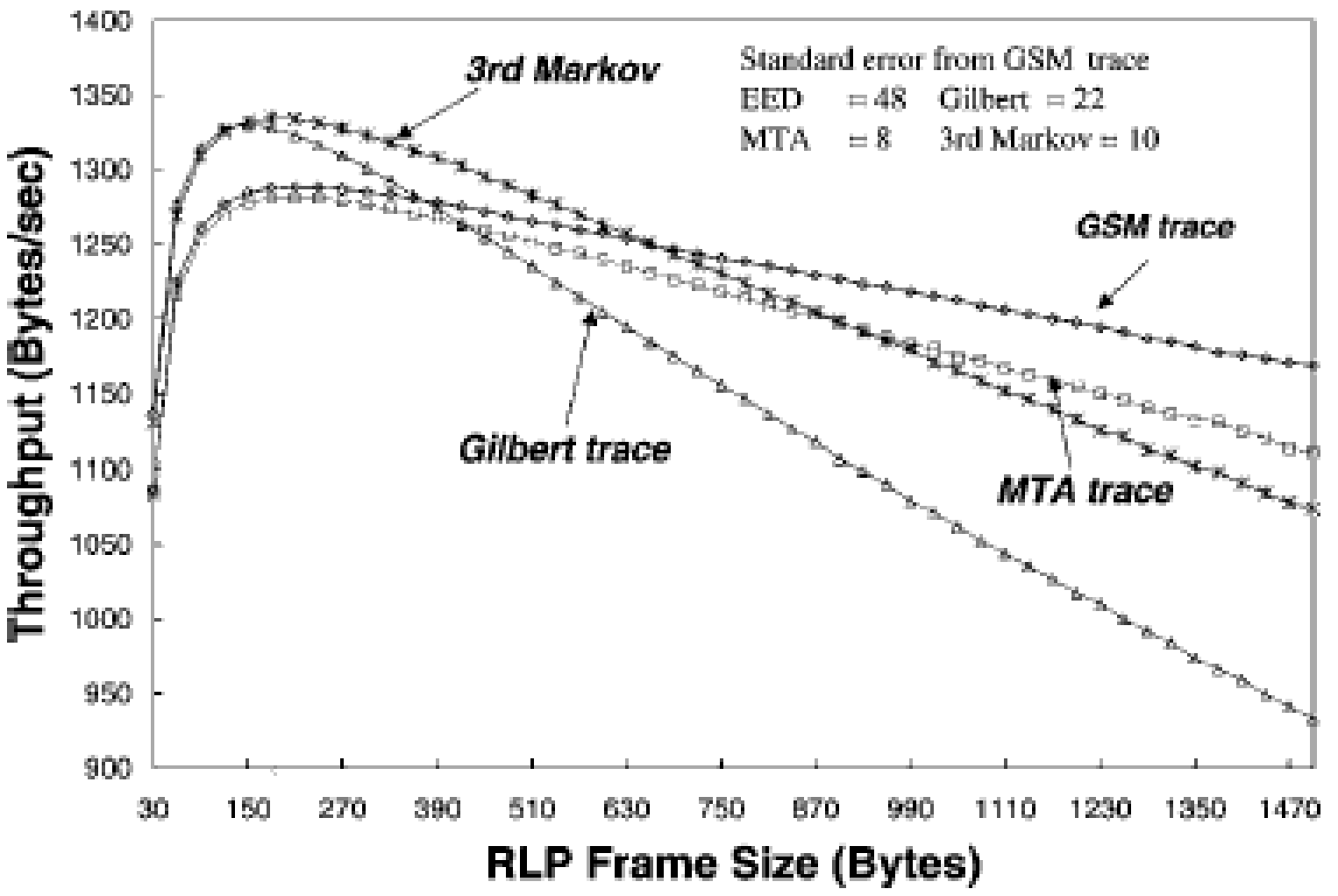

Fig. 4: Retrace analysis of the three methods over the GSM trace used in [8].
(EED method performance not shown)

The above analysis pertained to a particular channel used for transmission. Such a process can be done for different channels in the wireless spectrum and the quality of service provided by the channels can be thus evaluated. All the data can be made available to the central controller and sharing of the same would enable better decision making by the dynamic spectrum manager. To conclude, MTA model is a seminal work on channel state prediction and can be very well applied to model the wireless channel. However, one of the main distinctive points of this model (which could be considered a drawback as well) is that it does not have any kind of dependency with the distance between the nodes in the actual wireless channel, leaving its total character to the matrices that mark the system behaviour.

### B. HMM based Channel State Predictor

The method to be discussed here was presented by Chang-Hyun Park et al. in their paper "HMM based Channel Status predictor for Cognitive Radios" [10]. To implement the dynamic spectrum management, they had two components in their channel selector. A channel state predictor is part of this module and enables the channel selector to make its decision.

The authors proposed HMM algorithm that would study the data sequence available from the past communication history and could be used for channel state prediction. They implemented their HMM based predictor over MATLAB and their test data included periodic and aperiodic data. Periodic data was used to verify the performance of the predictor and then, the design was tested over aperiodic data set.

Their HMM model could be represented as having $\lambda = (A, B, \pi)$ and a sequence of observations $O = (o_1, o_2 \ldots o_T)$. The conditional probability $P(O|\lambda)$ was required to calculate the posterior probability (defined earlier under HMM) and usage of simple probabilistic analysis resulted in a very intensive computation method. So, an auxiliary variable called *forward variable* $\alpha_t(i)$, defined as probability of a partial observation sequence $(o_1, \ldots o_t)$ when analysis was terminated at time *t* and state *i,* was introduced. Another variable called *backward variable* $\beta_t(i)$ was defined as probability of partial observation sequence $(o_{t+1}, \ldots o_T)$ given that the current state is *i* and it was shown that $P(O|\lambda)$ could be expressed simply as summation of $\alpha_t(i)\beta_t(i)$ over all the states *i.* Now, posterior probability could be calculated. Lastly, using Baum-Welch algorithm to update the parameters of our starting model $\lambda = (A, B, \pi)$ through some *re-estimation formulae,* the posterior probability was maximized.

With the parameters of HMM available, the MATLAB simulator was ready for implementation. A pattern was loaded into the simulator and in the manner related above, the HMM trained and found its parameters and decided the most likely next state. Study of the success rates for different patterns loaded into the simulator showed that patterns with period more than 2 had less successful predictions whereas patterns that changed continuously or had period 1 showed good results.

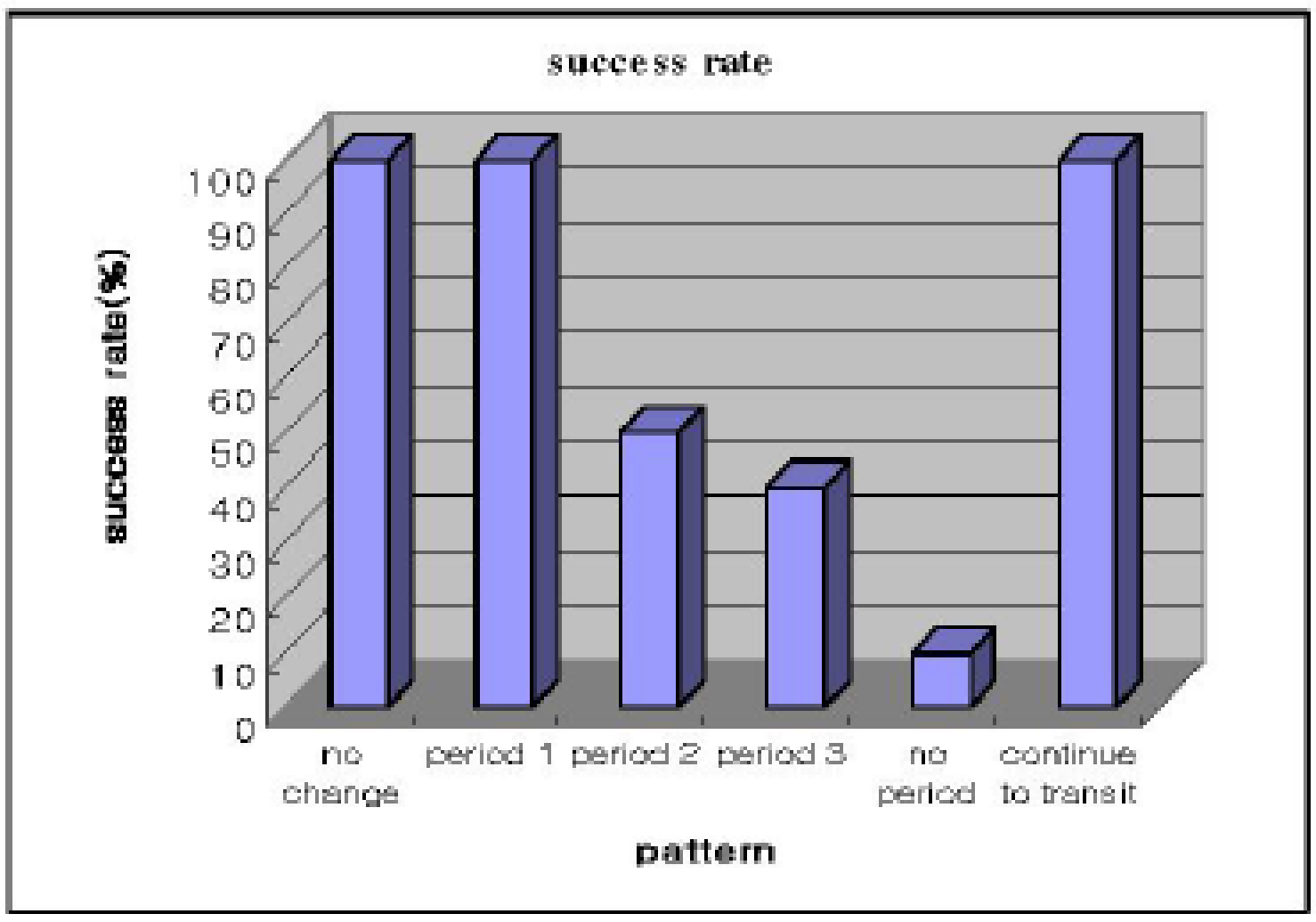

Fig. 5: Success rate of the HMM procedure.

Thus, it could be concluded that an algorithm like HMM which has applications in sequential pattern recognition, has disadvantages for periods more than 2. The simulator cannot find good parameters for the patterns and hence their false predictions. The pattern with period 1 had high success rate because the HMM parameters got trained quickly which could also be seen in their almost zero log-likelihood (a zero log-likelihood indicates perfect prediction). To achieve higher success for wide range of patterns, we may need to resort to other *heuristics*, may be like *genetic algorithm* (GA) but it has limitation in form of slow convergence. We may consider *particle swarm optimization* (PSO) or *ABC algorithms* that show better performance. A new heuristic called *Bat Algorithm* is proposed that makes use of the information in past solutions more effectively to generate better quality solutions faster, when compared to the other population-based optimization algorithms.

## IV. Conclusion

Cognitive Radio (CR) has thus emerged as one of the keys that can help address the inefficient usage of the radio spectrum, without requiring the allocation of new frequency bands by opening up the unused licensed frequencies to secondary users for opportunistic access. It intelligently adapts its spectrum usage according to the changes in radio environment as per some predefined objectives (availability, QoS etc.). So, efficient algorithms for learning based on experience and observation, and for decision-making are quite significant in this context.

There are many aspects of the CR functionality that are being currently researched into but we restrict the discussion to channel state estimation. We can extend *Conditional Random Field (CRF)*, a new family of discriminative statistical modelling method to this class of problem. It can better handle the known relationships between observations and construct consistent interpretations for it incorporates the properties of HMMs and the Maximum Entropy Model. In contrast to HMMs, CRFs are not tied to linear-sequence structure but can be arbitrarily structured [11]. So, the use of CRF is proposed for wireless channel state prediction based on the information fed by the receiver back to the transmitter.